\documentclass[10pt,conference]{IEEEtran}
\usepackage{graphicx}
\usepackage{epsfig}
\usepackage{url}
\usepackage{algorithm}
\usepackage{algorithmic}
\usepackage{amsmath}
\usepackage{float}
\usepackage{wrapfig}

\usepackage{soul}
\usepackage{color}
\usepackage{multirow}
\usepackage{graphicx}
\usepackage{subfigure}
\usepackage{comment}
\usepackage{balance}
\usepackage{epstopdf}
\usepackage{epsfig}
\usepackage{changepage}
\usepackage{xspace}
\usepackage{tabularx}
\usepackage{booktabs}
\usepackage{threeparttable}
\usepackage[utf8x]{inputenc}

\newcommand{\sysname}{\texttt{OTP-Lint}\xspace}

\newcommand{\tool}[1]{{\textsf{\small{#1}}}}
\newcommand{\func}[1]{{\texttt{{#1}}}}

\begin{document}

\title{Fine with ``1234''? An Analysis of SMS One-Time Password Randomness in Android Apps}
%

\author{
{\rm Siqi Ma$^{1}$},
{\rm Juanru Li$^{*2}$},
{\rm Hyoungshick Kim$^{3}$},
{\rm Elisa Bertino$^{4}$},
{\rm Surya Nepal$^{5}$},
{\rm Diethelm Ostry$^{5}$},
and
{\rm Cong Sun$^{6}$}\\

$^{1}$ The University of Queensland, slivia.ma@uq.edu.au \\
$^{2}$ Shanghai Jiao Tong University, jarod@sjtu.edu.cn \\
$^{3}$ Sungkyunwan University, hyoung@skku.edu \\
$^{4}$ Purdue University, bertino@purdue.edu \\
$^{5}$ Data61 CSIRO, \{surya.nepal, diet.ostry\}@data61.csiro.au \\
$^{6}$ Xidian University, suncong@xidian.edu.cn
}

\maketitle

\begin{abstract}
A fundamental premise of SMS One-Time Password (OTP) is that the used pseudo-random numbers (PRNs) are uniquely unpredictable for each login session. Hence, the process of generating PRNs is the most critical step in the OTP authentication. An improper implementation of the pseudo-random number generator (PRNG) will result in predictable or even static OTP values, making them vulnerable to potential attacks. In this paper, we present a vulnerability study against PRNGs implemented for Android apps. A key challenge is that PRNGs are typically implemented on the server-side, and thus the source code is not accessible. To resolve this issue, we build an analysis tool, \sysname, to assess implementations of the PRNGs in an automated manner without the source code requirement. Through reverse engineering, \sysname identifies the apps using SMS OTP and triggers each app's login functionality to retrieve OTP values. It further assesses the randomness of the OTP values to identify vulnerable PRNGs. By analyzing 6,431 commercially used Android apps downloaded from \tool{Google Play} and \tool{Tencent Myapp}, \sysname identified 399 vulnerable apps that generate predictable OTP values. Even worse, 194 vulnerable apps use the OTP authentication alone without any additional security mechanisms, leading to insecure authentication against guessing attacks and replay attacks. 
\end{abstract}



\section{Introduction}
\label{sec:intro}


SMS One-Time Password (OTP) is widely used for authentication and authorization in Android apps~\cite{mulliner2013sms}, which employs a uniquely generated pseudo-random number (PRN) for each login session to verify each user's identity. A pseudo-random number generator (PRNG) is commonly used to generate unpredictable OTP values.
Some cryptographically insecure randomness algorithms, such as Mersenne Twister (MT)~\cite{argyros2012forgot} and Linear Congruential Generator (LCG)~\cite{girault1988generalized}, have been used in practice. A PRNG using any of these insecure randomness algorithms would generate highly predictable OTP values. Even though the utilized randomness algorithm is secure, the generated PRNs may still be problematic if the algorithm is not implemented correctly (e.g., seeding the randomness algorithm by using a constant)~\cite{acar2017comparing,ma2019empirical}. 

Many studies~\cite{matsumoto1998mersenne, eichenauer1986non} have been proposed to analyze the security of pseudo-random number generating algorithms; however, these studies seldom analyze PRNG implementations in apps. The techniques proposed for assessing the PRNG implementations mainly focus on open-source systems (e.g., Linux~\cite{dodis2013security}, OpenSSL~\cite{kim2013predictability}). However, these techniques rely on code analysis and thus cannot be applied to analyze PRNGs of Android apps. This paper focuses on the following two goals: 1) exploring security vulnerabilities in SMS OTP values generated by Android apps; 2) gaining insights into potential implementation issues of PRNGs used by these vulnerable apps, without accessing the source code of the PRNGs.

Towards fulfilling our goals, we first study the algorithms and functions for generating PRNs to understand what types of vulnerabilities may occur in PRNGs. Next, based on the official RFC documents~\cite{rfc4086,saito2020tinymt32,m2011totp,m2005hotp} and research work~\cite{mascagni2004parameterizing,panneton2006improved,brown1994security}, we introduce three critical randomness rules: \textbf{Rule 1} -- Do not use a static OTP value; \textbf{Rule 2} -- Do not generate OTP values according to specific patterns; \textbf{Rule 3} -- Do not use a constant or predictable seed to initialize a randomness function (Section~\ref{sec:random}). If the OTP values generated for user authentication violate any of the randomness rules, those OTP values can be predicted, and thus the authentication scheme can eventually be cracked. 

We develop a novel analysis tool, \sysname, to assess the randomness of OTP values and analyze the potential implementation vulnerabilities of the corresponding PRNGs without having access to the PRNG source code. \sysname first identifies the login \textit{Activity} declared in each app using a fuzzing-inspired approach. It then recognizes those apps which use SMS OTP authentication through keyword matching. By locating the OTP login widgets, \sysname triggers the relevant app functionalities and sends OTP requests to the app server to retrieve OTP values. Finally, \sysname evaluates whether the gathered OTP values violate any of the three introduced randomness rules. A major challenge in the vulnerability analysis of OTP values is to determine which algorithm and function are used to generate PRNs and which parameters they are given as input.
To address this challenge, we collected PRNG sample codes written in diverse programming languages shared by app developers on \tool{GitHub}~\cite{github} and \tool{Stack Overflow}~\cite{stackoverflow} to learn the popular approaches for implementing PRNGs.

We used \sysname to analyze 6,431 real-world Android apps, downloaded from both \tool{Google Play} and \tool{Tencent MyApp} markets (1,000 from \tool{Google Play} and 5,431 from \tool{Tencent MyApp}). Out of these apps, \sysname successfully detected that 2,022 apps implemented SMS OTP login, and 399 (19.7\%) of these violated the defined randomness rules. Our results demonstrate that a significant number of Android apps would be at real risk of cyber-attacks exploiting such OTP login functions.


We believe that \sysname would help service providers and users test whether the OTP authentication in Android apps is securely implemented and highlight potential security issues without accessing PRNG source code. Our main \textbf{contributions} are as follows:

\begin{itemize}

\item Through an examination of the official RFC documents and research works, we provide insights into potential implementation issues in vulnerable PRNGs and propose three types of randomness rules that must be followed for implementing secure PRNGs. 

\item To the best of our knowledge, this is the first randomness study of OTP values generated by PRNGs used in Android apps. Without knowing the detailed implementations of PRNGs, we infer the potential vulnerabilities that might exist in the PRNGs through OTP value analysis.

\item We build a novel analysis tool, \sysname. By triggering OTP authentication in apps, \sysname simulates the most common vulnerable PRNG implementations for OTP randomness analysis and checks whether the generated OTP values are vulnerable.

\item We used \sysname to analyze 6,431 Android apps and detected 399 apps that produce predictable OTP values. Interestingly, 194 vulnerable apps use the OTP authentication alone without any additional security mechanisms, leading to insecure authentication against guessing attacks and replay attacks.

\end{itemize}

\section{Analysis of Randomness Weaknesses}
\label{sec:pre}
We discuss the widely used randomness algorithms and present the randomness functions in programming languages that can be used for Android apps.

\subsection{Randomness Algorithms}
\label{sub:algo}

\noindent
\textbf{Linear Congruential Generator (LCG).}
LCG is one of the most popularly used algorithms that generate a sequence of pseudo-random numbers (PRNs), using a discontinuous linear equation. LCG is defined by the recurrence relation $S_{n+1}=a \times S_n + C \mod m$. Starting with a seed, LCG repeatedly applies the recurrence relation to generate the subsequent PRNs. Such an algorithm is not cryptographically secure~\cite{krawczyk1992predict}. If a sufficient number of PRNs are gathered, an attacker can predict subsequent values.

\noindent
\textbf{Lagged Fibonacci generator (LFib).}
LFib is an improvement of the ``standard'' LCG, where PRNs are derived as a generalization of the Fibonacci sequence. Such a sequence is generated according to the recurrence relation $S_k=S_{n-j} \otimes S_{n-p} \mod m$ $(0<j<p)$~\cite{mascagni2004parameterizing}, where $\otimes$ is any binary function such as addition, subtraction, multiplication, or even the bitwise XOR. LFib thus requires an initial sequence and two seeds to begin. However, such algorithms using two seeds can be vulnerable to birthday attacks~\cite{girault1988generalized}. Alternatively, therefore, three seeds are recommended to be used according to the expression $S_k=S_{n-q} \otimes S_{n-j} \otimes S_{n-p} \mod m$ $(0<q<j<p)$. In addition, the initial sequence of LFib should contain at least one odd number; otherwise, the generated PRNs would be all even. However, even if these issues are addressed, LFib is still not cryptographically secure because it is represented as a linear recursion.


\noindent
\textbf{Mersenne Twister (MT).}
MT~\cite{matsumoto1998mersenne} is another popular algorithm. MT does not use any arithmetic operations (i.e., $+$, $\times$, $-$, $\div$) but is based on a group of permutation and tempering operations with shifts ($<<$ and $>>$), AND ($\&$), OR ($||$), and XOR($\mathbin{\oplus}$). Since MT is based on a linear recursion, it is not cryptographically secure~\cite{argyros2012forgot}. One can determine the internal state of the algorithm once a sufficiently long sub-sequence of outputs is observed.
The most common implementation of MT is MT19937, in which only 624 distinct outputs can be used to derive all the internal state variables of the PRNG~\cite{jagannatam2008mersenne}.


\noindent
\textbf{Well Equidistributed Long-period Linear (WELL).}
Similar to the MT algorithm, WELL is also a form of linear feedback shift register using simple bitwise operations. A seed value is required to start the generation process.  With only a slightly higher time cost, WELL obtains better equidistribution than MT~\cite{panneton2006improved}. WELL512, with a state size of 512 bits, is a widely used version. Its generated outputs are only selected within the restricted state instead of unbounded dynamic memory allocation. The period length of the generated PRNs is approximately $2^{512}$. At first glance, the use of WELL512 is sufficiently secure for OTP generation. However, when the same seed is used, WELL512 also becomes vulnerable~\cite{well}.


\subsection{Randomness Functions}
Most programming languages provide the functions for generating PRNs by default. Instead of analyzing all of them, we only consider the most common ones --- C/C++, Python, PHP, Java, and JavaScript~\cite{bissyande2013popularity}. We introduce the randomness functions that are frequently used to generate PRNs and discuss each function's security issues.

\subsubsection{C/C++} Many PRNG functions are provided in the C library. Two primary functions are listed below.

\noindent
\textbf{rand()}: This function is a C standard built-in generator. To ensure that the sequence of PRNs is unpredictable, \func{srand($\cdotp$)} is called to initialize the PRNG with a seed value beforehand. The algorithm of \func{rand()} is adapted from the BASIC PRNG algorithm, and simple operations such as arithmetic (e.g., $\times$) and bitwise operations (e.g., $\&$) are involved.
If a static seed is used, the \func{rand()} generates the same stream of PRNs. Hence, in order to produce an unpredictable sequence of PRNs, the initialization should not be a constant value, but a pseudo-random value instead. Nonetheless, \func{rand()} uses the LCG algorithm without adding any entropy to the generator, which is not cryptographically secure. Although the LCG parameters are unknown, the attacker can easily identify the parameters when consecutive outputs are collected.  

\noindent
\textbf{rand\_s()}:
This is a secure alternative for \func{rand()}. It generates cryptographically secure PRNs depending on the operating system. It is not affected by the seed produced by \func{srand()}; it also does not affect the pseudo-random number sequence used by \func{rand()}.
It is essential to mention that this function only works on Windows XP and its later versions.

\subsubsection{Python} Python uses the MT algorithm as the core generator and leverages the standard MT implementation (i.e., MT19937). Hence, the randomness functions in Python are unsuitable for cryptographic purposes.

\noindent
\textbf{random.randrange(start, stop)}: This function generates a pseudo-random integer within a range of [start, stop]. By default, \func{start} is defined as zero. 
    
\noindent
\textbf{numpy.random.rand($d_0, d_1,\ldots, d_n$)}: This function takes as input the dimensions of the array to be created. It then creates the array and fills it with PRNs from a uniform distribution over [0, 1).

\subsubsection{PHP} PHP is the most popular programming language for server-side scripting. Three main functions are discussed.

\noindent
\textbf{lcg\_value()}: This function is used in numerous places within the Zend engine code as an internal function.
    This function leverages two LCGs to get better quality PRNs. A default seeding algorithm uses time and process id inputs to calculate a value to seed both LCGs. 
    The function implementation is proved to be cryptographically insecure as one can determine the generated values by only using consecutive outputs~\cite{argyros2012forgot}.
    
\noindent
\textbf{rand(min, max)}: This function generates a pseudo-random integer. By default, it falls back to \func{rand()} supported by C with the range [min, max] is [0, $2^{31}-1$]. The implementation of this PRNG generates a default seed by taking the current timestamp and the value generated by \func{lcg\_value()} as input.
    Alternatively, users can call \func{srand($\cdotp$)} to set the seed externally. When a constant or predictable seed is chosen, this PRNG becomes insecure.
    
\noindent
\textbf{mt\_rand()}: This function is implemented using the MT algorithm. By implementing MT19937, namely, the standard implementation of MT, this function is not cryptographically secure because the algorithm's internal states can be observed when 624 successive outputs are gathered.

\subsubsection{Java} There are three primary functions for PRNG implementations in the Java development kit (JDK).  

\noindent
\textbf{random()}: This function is included in the class \func{java.util.Random} for generating a stream of PRNs with positive signs and restricted within [0.0, 1.0]. This function is created based on Knuth's subtractive algorithm~\cite{hanna2010emperor}, that is, an internal PRNs repeated cycle length is fixed. In addition, this class utilizes LCG with a 48-bit static seed. Without any entropy added to the generator, this function is not cryptographically secure because two pseudo-random sequences created by the same seed are the same. 
When the PRNG is unknown, the PRNG parameters can be calculated when $2^{32}$ consecutive outputs are known.

\noindent
\textbf{Math.random()}: This function is included in the \func{java.util} package. Without giving any seed, a \func{java.util.Random} object is created when \func{Math.random()} is called; hence the PRNs generated by \func{Math.random()} depend on \func{random()}. It is also cryptographically insecure.

\noindent
\textbf{SecureRandom()}: This function is included in the Java class \func{java.security.SecureRandom} to produce secure PRNs. By using SHA-1 as part of the random algorithm, the generated PRNs are hashed; hence \func{SecureRandom()} provides a strong PRNG implementation to ensure a non-deterministic output.

In Java, it is thus critical to use the randomness function \func{SecureRandom} in \func{java.security.SecureRandom} class instead of using \func{random()} or \func{Math.Random()} to generate PRNs for security reasons.

\subsubsection{JavaScript.} 
JavaScript also provides \func{Math.Random()}, which is insecurely implemented in the same manner as Java. However, several cryptographically secure randomness functions are also supported. A web cryptographic function \func{window.crypto.getRandomValues} with a high entropy seed is recommended. Furthermore, a secure crypto library (\tool{SJCL})\footnote{SJCL: https://crypto.stanford.edu/sjcl/} is also provided.

\section{Randomness Rules for One-Time Password}
\label{sec:random}
While weaknesses and recommended implementations of the randomness algorithms and functions are precisely described, we are curious whether the PRNGs implemented by developers achieve the expected security level. Therefore, we introduce three generic types of randomness rules for OTP. The sequence of the gathered OTP values that violate any of these rules indicates that the OPT values can be predicted with a significant probability. Note that we assume the network channel is secure when OTP values are transmitted.

\noindent
\subsubsection{Rule 1. Do not use a static OTP value~\cite{ma2019empirical}}
It forbids using a static value for all login sessions of different accounts, which violates the requirement of OTP randomness. Consequently, the security of the OTP authentication scheme is not guaranteed once the OTP value is exposed to attackers. 

\noindent    
\textbf{Threat.} 
When a static OTP value is used, the OTP authentication is vulnerable to replay attacks. If the length of the OTP value is insufficient~\cite{m2011totp}, the OTP is guessable through brute force attacks.

\subsubsection{Rule 2. Do not generate OTP values according to specific patterns~\cite{m2011totp}~\cite{m2005hotp}}
OTP values should be unpredictable. However, some real-world randomness functions generate OTP values in specific patterns, which can be statistically observed.
We discuss the three sub-rules below.

\noindent
\emph{Rule 2-1. Do not generate a repeated sequence of OTP values:}
It states that the PRNG should not generate PRNs with a fixed period length. Unsurprisingly, if OTP values are periodically repeated, attackers can easily predict the OTP values when the number of generated OTP values is larger than the period size. 

\noindent
\emph{Rule 2-2. Do not repeat each distinct OTP value $n$ times:}
It demonstrates that each OTP value should not be used repeatedly for the $n$ consecutive login sessions.

\noindent
\emph{Rule 2-3. Do not generate OTP values with predictable binary representations:}
It states that the PRNs should be pseudo-random in any format. Even though the generated OTP values in the decimal format seem unpredictable, they may have some specific patterns in the binary format. For example, the parity of OTP values has a specific pattern such as all evens or $(\emph{odd}, \emph{even}, \emph{odd}, \emph{even}, \dots)$ parity. In such cases, the possible space of the possible OTP values can be reduced significantly.

\noindent
\textbf{Threat.}
OTP authentication is vulnerable to replay attacks when the generation pattern of OTP values is exploited.
The attacker first collects a certain number of the login communication packets. Without retrieving the plaintext of the OTP values, the attacker can send the corresponding packet to the server consistent with the generation pattern.

\subsubsection{Rule 3. Do not use a constant or predictable seed to initialize a randomness function~\cite{prng}}
It states that the randomness functions should not be seeded with a constant or predictable seed. When the seed is a constant value, the attacker can duplicate the sequence of PRNs when the seed is guessed.
For example, when using \func{srand(1)} to seed \func{rand()}, \func{rand()} always outputs the same sequence because the initialization status is determined. Note, when a dynamically changed seed is used for the randomness function, the OTP authentication can still be insecure if the seed is predictable (e.g., use of a timestamp)~\cite{boyar1989inferring}.

\noindent
\textbf{Threat.}
When a randomness function (e.g., \func{rand()}) is seeded by a constant or a predictable value, the attacker can test the possible seeds through brute force attacks and infer the following OTP values.

\section{Challenges}
\label{sec:challenge}
Three challenges need to be addressed while analyzing the real-world apps to identify the implementations violating the randomness rules introduced in Section~\ref{sec:random}.

\noindent
\textbf{Challenge 1: How can we determine the selected randomness algorithm and the PRNG implementation resided on the server-side?} App developers typically select a particular randomness algorithm to generate OTP values. However, general program analysis techniques such as program slicing~\cite{weiser1984program} cannot be used here. We cannot access the OTP generation implementation because it resides on the server-side. Diverse PRNG implementation options (written in many different programming languages) make it harder to decide which PRNG is specifically used for the OTP functionality.

\noindent
\textbf{Challenge 2: How many OTP values should be gathered to infer potential patterns in an OTP sequence?}
A large number of OTP values are needed to determine whether any pattern exists in the OTP sequence. Therefore, multiple login attempts should be made. However, it is time-consuming or often not allowed to gather a massive number of OTP values through login attempts. Therefore, we need to minimize the number of login attempts as much as possible.

\noindent
\textbf{Challenge 3: How can we collect OTP values for our experiments without affecting the OTP server?}
We can perform the login process repeatedly to collect a sufficient number of OTP values. However, sending a large number of login requests can interfere with the OTP server's normal operations. More worryingly, it can raise ethical issues.

To address these challenges, we propose three approaches: 
1) We analyze the existing PRNG implementations written in diverse programming languages and then abstract the common implementations in Android apps.
2) Based on the analysis of existing PRNG implementations, we obtain vulnerable codes in each code snippet. If a predictable pattern is found, we then determine how many OTP values are sufficient to infer the following PRNs.
3) We set a maximum number of login attempts for collecting OTP values because we need to stop the OTP data collection process when we fail to identify a specific pattern from a sequence of OTP values. Moreover, since some apps limit the number of login attempts per day (e.g., 20 per day), it would be time-consuming to collect the number of OTP values larger than such a maximum number of login attempts. Therefore, we empirically set the maximum login attempts as 1,000 by considering both practicality and efficiency.

\section{Overview of \sysname}
\label{sec:sys}
\begin{figure}[!ht]
    \centering
    \setlength{\abovecaptionskip}{-0.2cm}
    \setlength{\belowcaptionskip}{-20cm}
    \includegraphics[width=0.7\linewidth]{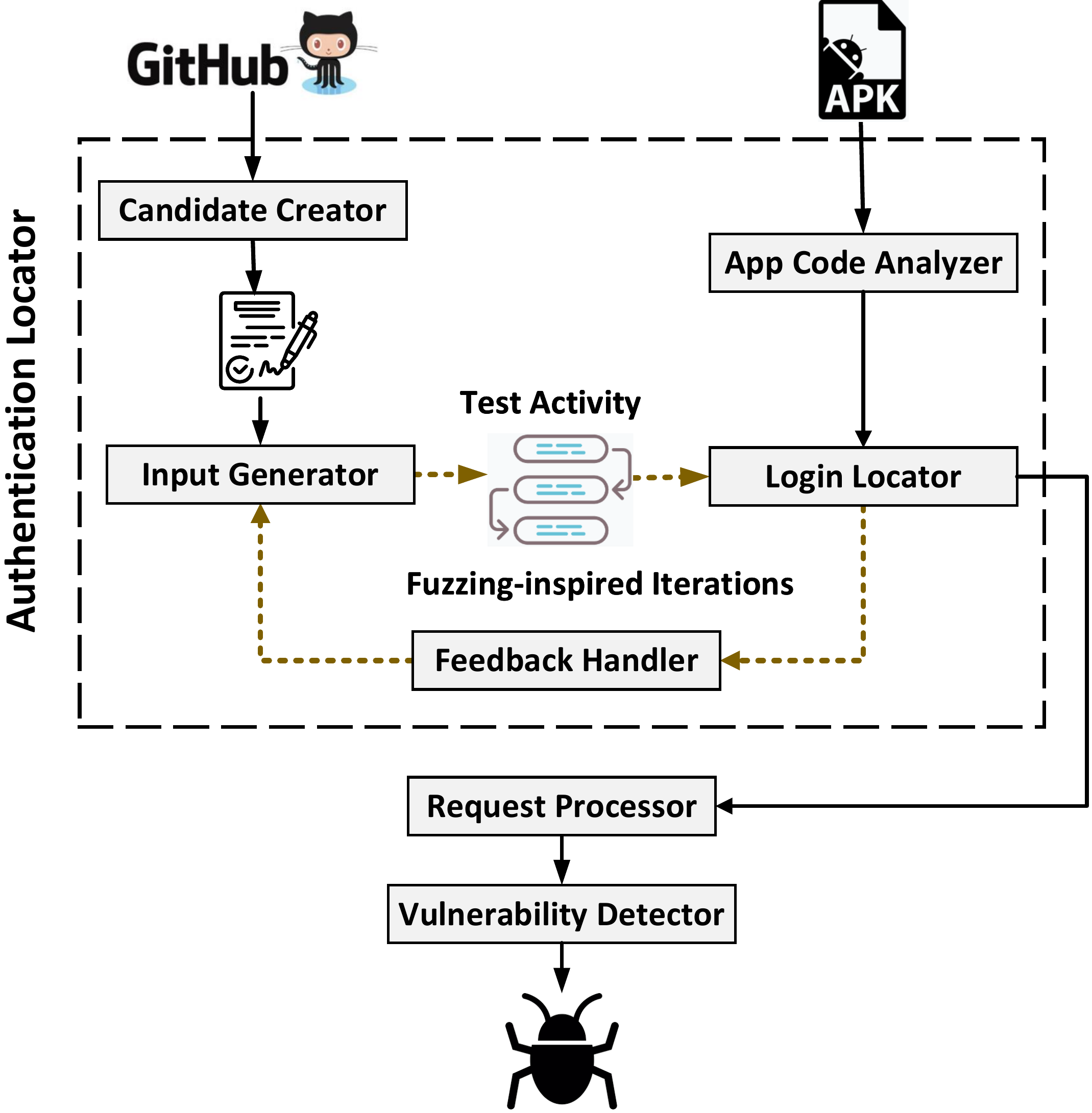}
    \caption{Workflow of \sysname.}
    \label{fig:sys}
\end{figure}

To investigate the randomness of OTP values, we build \sysname to collect a sufficient number of OTP values from a given Android app using the SMS OTP authentication. \sysname then checks whether randomness rules listed in Section~\ref{sec:random} are violated.
Figure~\ref{fig:sys} shows the workflow of \sysname, which includes three components: \emph{Authentication Locator}, \emph{Request Processor}, and \emph{Vulnerability Detector}.

\subsection{Authentication Locator}
\label{sub:locator}
In order to trigger login requests for OTP collection, \sysname first identifies whether there are any login \textit{Activities} implemented in apps.
The variety of login \textit{Activities}, which are named differently, makes it challenging to identify login \textit{Activities} by simply using keywords matching~\cite{wang2019nlp}~\cite{ma2019empirical}. The customized functions are generally declared by using abbreviations and informal terms (e.g., \func{AccountAct}, \func{AuthAccount}), which are difficult to recognize. 

For two functions using similar keywords, these functions' semantics might be different if the order of these keywords in the two names is different.
As an example consider the functions \func{SMSLoginAct} and \func{LoginSMS}. They have different semantics although both of them utilize the keywords \func{SMS} and \func{login}. The \func{SMSLoginAct} function represents a SMS login \textit{Activity}, whereas function \func{LoginSMS} requires the OTP value for identity verification.
Therefore, only matching the keywords makes the \textit{Activities} identification inaccurate.

In order to achieve an accurate login \textit{Activity} identification, \sysname applies a fuzzing-inspired approach. Fuzzing~\cite{sutton2007fuzzing} is an automatic test generation and execution to find security vulnerabilities. We aim to combine fuzzing with program analysis to address the issues of inconsistent function naming and ambiguous function semantics. Instead of matching keywords, \sysname relies on code dependencies to understand the purpose of each function. In particular, it creates a test Activity consisting of dependencies to locate login \textit{Activities}. According to the previous test executions' feedback, \sysname dynamically optimizes the dependencies in the test Activity for the next round of \textit{Activity} identification.

The authentication locator executes five steps:
1) a \textbf{candidate creator} generates candidate samples;
2) an \textbf{app code analyzer} analyzes the code of the target app and extracts dependencies; 3) an  \textbf{input generator} creates a test Activity;
4) a \textbf{login locator} identifies login \textit{Activities} from the real-world Android apps; and
5) a \textbf{feedback handler} optimizes
the test Activity when the login \textit{Activity} is not identified in an app.
\sysname executes steps 3--5 iteratively until the number of iterations exceeds a threshold, which indicates that there would likely be no login \textit{Activity} implemented in the apps. Referring to the previous fuzzing approaches~\cite{zhang2020ethploit}~\cite{godefroid2008automated}, we set the iteration threshold as 1,000 by considering both effectiveness and efficiency. 

\subsubsection{Candidate Creator}
Candidate samples are a group of sample codes used as references when \sysname is creating the test \textit{Activity} for further login \textit{Activity} identification.
However, it is difficult to determine whether a piece of code is relevant to user authentication. Hence, \sysname obtains the authentication code from GitHub repositories.

Specifically, \sysname first crawls all the repositories by searching for the keyword \func{authentication} and selects those whose \tool{read.md} and the title contain the keyword. 
As login \textit{Activities} in Android apps are typically written in Java and the invoked functions are Java methods, \sysname initially takes as input the GitHub repositories for authentication~\cite{auth}. As Android apps are generally written in Java, we only consider the repositories in Java. In total, 9,134 GitHub repositories were analyzed to construct a candidate list. 

Instead of analyzing the entire project, \sysname only extracts the code snippets that are relevant to authentication. Each candidate is represented according to the format of $C=(tx_1, tx_2, \dots, tx_n)$, where $C$ is the authentication method name and $(tx_1, tx_2, \dots, tx_n)$ are the invoked functions.
Two steps are proceeded to construct each candidate:

\noindent
\textbf{Step 1: Method Determination.} 
To identify authentication code snippets, we follow the same natural language processing technique used by \tool{NLP-EYE}~\cite{wang2019nlp}. First, we manually recognize the login-related classes from the GitHub repositories to construct a reference set. Then we use all posts on Stack Overflow\footnote{Stack Overflow is the most popular web service to programmers for discussing technical issues, in the form of Question and Answers.} (\url{https://stackoverflow.com/}) to generate a code corpus. Given each method name in a project, \sysname compares it with the words in the reference set and uses the code corpus to measure the semantic similarity between the method name and the word in the reference set. If two words are regarded as similar, then the method is labeled as an authentication method.

\noindent
\textbf{Step 2: Dependency Construction.}
After identifying the authentication method, \sysname derives the in-method and external-method correlations by conducting the intra- and inter-dependency analyses. To extract the in-method correlations, \sysname constructs an intra-dependency graph to extract in-method correlations by exploring the invoked functions. Nodes in the intra-dependency graph and the inter-dependency graph are the invoked functions, and each directed edge represents a dependency from a caller function to a callee function. By creating the inter-dependency graph, \sysname obtains the external-method correlations. It locates where the customized function is declared and constructs an intra-dependency graph for the customized function. Through the customized function, its intra-dependency graph is connected with the graph of the authentication method.
  
Since some Java functions that execute common behaviours (e.g., \func{nextLine()}, \func{toString()}) and exception handling functions (e.g., \func{printStackTrace}) might affect the detection precision, \sysname further removes these redundant functions.

\subsubsection{App Code Analyzer}
\sysname takes each Android app as input and proceeds two steps: \emph{Decompilation} and \emph{Dependency Construction}.
First, \sysname uses \tool{JEB} Android decompiler~\cite{jeb} to decompile each app into Java source code.
Similar to the candidate creator's dependency construction step, \sysname constructs the intra- and inter-dependency graphs to extract the in-method and external-method correlations.
Since we do not know which methods are relevant to authentication, \sysname targets the entire app code to conduct the dependency analysis. Again, the redundant functions are excluded in the generated dependency graphs.

\subsubsection{Input Generator}
\sysname creates a \emph{test Activity} that can be adapted to search for the authentication login \textit{Activity} in each app. Instead of manually defining the test Activity, which would be time-consuming and inaccurate, the input generator constructs and optimizes the test Activity by using a fuzzing-inspired approach.

A test Activity consists of the functions that are commonly invoked for authentication login, represented according to the notation $A=(fc_1, fc_2, \dots, fc_x)$, where $A$ is a login Activity represented by a two-tuple ($Name$, $Argument Names$) and $(fc_1, fc_2, \dots, fc_x)$ are a sequence of function names that must be invoked for authentication. \sysname optimizes the test Activity used for the previous iteration to generate the next test Activity. It takes the following two steps. 
Note that \sysname randomly selects a candidate from the candidate list as initial input for the input generator.

\noindent
\textbf{Step 1: Activity Selection.}
\sysname reconstructs the test Activity $A$ by referring to the candidate samples. By comparing $A$ with the candidate samples, \sysname finds the candidate $C_{sim}$ that is the most similar to $A$ and then selects the functions in $C_{sim}$ to replace the inappropriate functions in $A$.
Such a reconstruction follows two principles:
\begin{itemize}
\item Candidate Selection -- Since every single word generally covers one functionality to ensure code readability~\cite{butler2011mining}, we assume that two methods with similar method names may fulfill the same functionality.
Therefore, \sysname pinpoints $C_{sim}$ by extracting the longest common substring (LCS)~\cite{flouri2015longest}.
Considering the length of the LCS as the similarity score between $A$ and each candidate, \sysname selects $C_{sim}$ with the highest similarity score.

\item Probability distribution -- Although $A$ and $C_{sim}$ are assumed to fulfill a similar functionality, the invoked functions in $A$ and $C_i$ might be different. Hence, \sysname randomly selects only one function $tx_j$ in $C_i$ to replace the function in $A$ at the same position during each iteration.
\end{itemize}

\noindent
\textbf{Step 2: Argument Creation.} After selecting the functions in the test Activity selection, \sysname fills in the required arguments for the test Activity.
With the matched candidate $C_i$, \sysname identifies the arguments that do not exist in $A$ and then inserts the corresponding argument names.

\subsubsection{Login Locator}
Given the test Activity and the dependency graphs of each Android app, \sysname locates the login \textit{Activity} via \emph{Activity Searching}. Since we only consider the \textit{Activities} defined in each app, \sysname selects the dependency graphs that illustrate the \textit{Activities}. Each dependency graph will be compared against the test Activity to verify whether there is any dependency graph that contains a subgraph, which is isomorphic to the test Activity. A graph is considered relevant to the login \textit{Activity} if a subgraph is identified; otherwise, \sysname selects the next dependency graph.

When all the dependency graphs are analyzed, and none of them are defined as login \textit{Activity}, \sysname continues the following steps to optimize the test Activity.
Alternatively, \sysname reports the identified login \textit{Activity} for further analysis.

\subsubsection{Feedback Handler}
To optimize the test Activity, \sysname analyzes the similarity between the test Activity and each dependency graph. Since the test Activity is composed of functions named variously by different developers, \sysname compares the test Activity $A$ and the dependency graph $G=(g_1, g_2, \dots, g_m)$ function by function. 

\sysname conducts a transformed pairwise comparison~\cite{ma2019finding} to compare each $fc$ in $A$ with all $g$ in $G$. It first computes the similarity score of two functions by using the length of LCS and then constructs a similarity set, which consists of the similarity scores of $fc$. 
While comparing $fc_i$ and $g_j$, \sysname only considers $fc_i$ and $g_j$ as similar if and only if the length of LCS is higher than a threshold $LCS_{thresh}$. Otherwise, \sysname follows the dependency in $G$ to compare $fc_i$ with the rest functions, i.e., $[g_{j + 1}, g_m]$.

Among the similarity sets, \sysname chooses the set with the highest total score and replaces the corresponding function with the lowest similarity score in the next iteration.

\subsection{Request Processor}
\label{sub:request}
After locating the login \textit{Activity} in each app, \sysname follows the steps used by \tool{AUTH-EYE} to identify the app with the SMS OTP authentication scheme and then triggers the login request button to retrieve OTP values from the app server.

\noindent
\textbf{Step 1: SMS OTP Location. }
For each SMS OTP authentication scheme, the specific widgets (i.e., EditText and Button) are required. Therefore, \sysname identifies the SMS OTP authentication scheme by recognizing whether there exist relevant widgets. Since widgets used in apps are named typically, which is easier for users to recognize, \sysname performs keyword matching directly to identify the necessary widgets. 
We created a keyword list containing the keywords such as ``sms'' and ``mobilephone''.
To extract the widget information (i.e., type, text, orientation, and layout), \sysname utilizes \tool{UI Automator} to parse the XML layout files.
Then it matches the keywords with the text in the field of \func{android:text} to identify an SMS OTP login. As a result, a list of apps that implement SMS OTP authentication is retrieved.

\noindent
\textbf{Step 2: OTP Extraction.}
In order to send OTP requests automatically, \sysname executes \tool{Monkey}\footnote{UI/Application Exerciser Monkey: It acts as a stress test on the developed app, downloaded from https://developer.android.com/studio/test/monkey} to fill in the mobile phone number and trigger the button to start a login attempt.
However, \tool{Monkey} only generates pseudo-random streams of user events without locating where the corresponding widgets are. In terms of the collected widget information (e.g., layout, type), \sysname uses \tool{UI/Application Exerciser} to locate EditText and Button and calls the \func{dispatchString} method to enter a valid phone number. Because the Android phone involved in the experiment is rooted, \sysname obtains the returned SMS messages from the dataset \func{/data/data/android.providers.telephony/\\databases/mmssms.db}. Finally, it parses the SMS messages to extract the OTP values.

\subsection{Vulnerability Detector}
\label{sub:detector}
\sysname evaluates whether randomness of OTP values produced by real-world Android apps violates the randomness rules summarized in Section~\ref{sec:random}. We set the waiting time interval between two login requests to one minute.

\subsubsection{Rule 1: Do not use a static OTP value}
To identify whether a static OTP value is used, \sysname first requests five OTP values and checks whether they are the same. If so, \sysname then requests 15 OTP values to check whether the value changed\footnote{In many apps, the maximum number of login attempts allowed per day is 20.}. \sysname labels an app generating static OTP values when the 20 OTP values are the same. When an app renews the OTP value within the 20 times of login requests, such a vulnerable situation is discussed in Rule 2-2. In this paper, we only retrieve OTP values without consuming them. However, we observed that some apps would only generate a new OTP value when the previous value is consumed. Such a scheme is also insecure because attackers have time to launch attacks before the OTP value is consumed. 

\subsubsection{Rule 2: Do not generate OTP values according to specific patterns}
Without the knowledge about the PRNG implementation on the server-side, we downloaded the sample codes of the PRNG implementations shared on GitHub~\cite{github} and Stack Overflow~\cite{stackoverflow} as references. Then we identify a range of implementations that violate the randomness rules.

\noindent   
\textbf{Rule 2-1: Do not generate a repeated sequence of OTP values.}
Referring to the PRNG functions introduced in Section~\ref{sec:pre}, we select the vulnerable implementations and identify the fixed repeat length of the PRNG. According to the sequence length, \sysname sends a sufficient number of login requests to check whether the OTP values are repeated after a certain number of values are generated.
In particular, when a PRNG only generates $N$ distinct PRNs (i.e., the fixed repeat length of the PRNG is $N$), \sysname sends $2\times N$ OTP requests and then compares the sequences $\{1_{st}, 2_{nd}, \dots, N_{th}\}$ and $\{(N+1)_{th}, (N+2)_{th}, \dots, {2\times N}_{th}\}$.
If two sequences are identical, \sysname labels the PRNG as insecure.

\noindent
\textbf{Rule 2-2: Do not repeat each distinct OTP value $n$ times.}
\sysname examines generated OTP values to determine whether an OTP value is repeated consecutively (i.e., for $n$ consecutively login sessions, where $n=2,3,\dots$). The PRNG is labeled as insecure when an OTP value is repeated $n$ times.

\noindent
\textbf{Rule 2-3: Do not generate OTP values with predictable binary representations.}
Some OTP values seem to be pseudo-random in their decimal formats. While analyzing the sample code, we found that some developers create their PRNGs using shift operations instead of invoking the corresponding APIs. Thus, \sysname converts the decimal format into the binary format to check whether there is any pattern in generated PRNs.

Since various shift operations can be utilized, it is challenging to determine generally about what operations are carried out. We only focus on the simple shift operations (i.e., in clockwise/anticlockwise). 
First, \sysname obtains an OTP value and transforms the value into its binary format. According to the number of digits included in the binary value, \sysname requests the corresponding number of OTP values and retrieves their binary values.
By comparing the $i^{th}$ and the $(i+1)^{th}$ values, \sysname identifies whether the $i^{th}$ value can be transformed to the $(i+1)^{th}$ value through digit shifting (either forward or backward). 
Additionally, we also observed that the digits in some sequences do not shift iteratively, but a new digit (``1'' or ``0'') is inserted at the end of the sequence. Therefore, \sysname only analyzes how the digits of the first binary value shift instead of considering the newly inserted digits.
If a shift operation is recognized, the PRNG is labeled as insecure.

For the OTP sequences whose vulnerable patterns are not identified, \sysname checks the last digit of each OTP value. If the last digit is `0', \sysname labels it as odd; otherwise, it labels it as even. To determine whether a stream of pseudo-random values is parity-guessable, we send the login requests 20 times. If the received 20 OTP values appear by following a specific parity pattern, \sysname labels the PRNG as insecure.

\subsubsection{Rule 3: Do not use a constant or predictable seed to initialize a randomness function}
Because the same stream of PRNs can be generated when a static value is used as the seed of PRNGs, we manually inspect the PRNG sample code to learn what static values are frequently used. As a result, the \func{rand()} functions in C/C++ and PHP are commonly seeded by \func{srand(1)}; thus we simulate the PRNG by using \func{srand(1)}.
First, \sysname sends 1,000 login requests to retrieve 1,000 OTP values by considering both ethnic and efficiency issues (refer to Section~\ref{sec:challenge}). 
It then analyzes the length of the OTP values and executes the PRNG simulation to generate 50 PRNs in the same length.
If the sequence of the PRNs is a subsequence of the OTP sequence, \sysname labels the app PRNG as insecure.

Apart from that, we found that many posts recommend developers to seed randomness functions by using a timestamp (i.e., srand(time(0))). 
Therefore, we create a PRNG simulation by using a flexible seed. To ensure that the same timestamp is used for the simulated PRNG and the PRNG in an app, \sysname executes the simulated PRNG and simultaneously sends the login requests. Then it determines whether the values generated by the simulated PRNG and the app PRNG are the same.

\section{Evaluation}
\label{sec:eva}
Our evaluation has two goals. The first is to conduct a large scale randomness analysis on OTP values. The second is to inspect the analysis result manually to understand the detail implementations of these vulnerable PRNGs.

\subsection{Dataset}
We downloaded 6,431 top list apps from both \tool{GooglePlay} and \tool{Tencent MyApp} market (1,000 from \tool{Google Play} and 5,431 from \tool{Tencent MyApp})\footnote{We found that the apps published on GooglePlay barely use OTP authentication; thus we focus more on the Tencent MyApp market.}. When an app is existed in both app stores, we assumed the implementations of the both apps are the same and removed the one from the \tool{Tencent MyApp} dataset. 
The apps are selected from 21 categories\footnote{
Categories: Communication, Education, Health \& Fitness, Medical, Books \& Reference, Photography, Productivity, Video Players \& Editors, Travel \& Local, Map \& Navigation, Entertainment, Lifestyle, Shopping, Tool, News \& Magazine, Personalization, Productivity, Social, Beauty, Finance, Music \& Audio, and Parenting} and the 300 top listed apps in each category are obtained.

\subsection{OTP Login Activity}
\label{sec: OTP Login Activity}

\sysname successfully analyzed 4,015 out of 6,431 apps. The failed cases are discussed in detail in Section~\ref{sub:limitation}.
Through the fuzzing-inspired approach, \sysname found 3,657 apps with login \textit{Activities}. Among these apps, \sysname further identified 2,022 (55.29\%) apps with SMS OTP authentication; thus, the following experiments were conducted on these 2,022 apps.
  
By manually inspecting the apps with SMS OTP authentication, we found that only 214 were from \tool{GooglePlay}, and all of them utilize two-factor authentication (i.e., using both password-based authentication and OTP authentication). The rest 1,808 apps are from \tool{Tencent MyApp}; of these apps 1,068 (59.1\%) implement two-factor authentication, and 740 (40.9\%) only use a single OTP authentication scheme. Since we only analyzed the randomness of OTP values, the parts related to password authentication, i.e., username and password, were filled in manually. \sysname then only executed the OTP login procedure. For the experiments, we registered an account manually in advance for the each tested app.

\subsection{Results}

We report the randomness analysis result in Table~\ref{table:random}. 
In total, 399 (19.7\%) apps out of the 2,022 apps exhibit of violating the randomness rules we have introduced.

\begin{table}[!htp]
\centering
\setlength{\abovecaptionskip}{-1pt}
\setlength{\belowcaptionskip}{-5cm}
\caption{Violations of randomness rules.}

\begin{tabular}{l|l|c}
\toprule[1.5pt]
\multicolumn{2}{l|}{Violated Rules} & \# of Apps\tabularnewline
\midrule[1pt]
\multicolumn{2}{l|}{Rule 1} & 41 \tabularnewline
\hline 
\multirow{4}{*}{Rule 2} & Rule 2-1 & 162\tabularnewline
\cline{2-3} 
 & Rule 2-2 & 67\tabularnewline
\cline{2-3} 
 & Rule 2-3 & 125\tabularnewline
\cline{2-3} 
 & Total & 354 \tabularnewline
\hline 
\multicolumn{2}{l|}{Rule 3} & 4\tabularnewline
\hline
\multicolumn{2}{l|}{Total} & 399 (out of 2,022) \tabularnewline
\bottomrule[1.5pt]
\end{tabular}
\label{table:random}
\end{table}

\subsubsection{Rule 1: Do not use a static OTP value}
\sysname detected 41 (10.3\%) apps that produce OTP values violating Rule 1. That is, those apps' PRNGs use a static OTP value instead of generating dynamically modified OTP values for different login sessions.
Since only one account was created for each app and the received OTP values were not consumed, we were interested in finding out: 1) whether the OTP value changes after it is consumed; and 2) whether the apps produce the same OTP value for all accounts.

We then conducted two experiments: 1) consuming each OTP value after it is received; 2) registering an additional account and then sending the login requests through the same procedure. The first experiment showed that 20 apps kept returning the same OTP values, although the values were consumed, which indicates that OTP authentication in these apps is mistakenly implemented as password authentication. Due to the short length (typically in four to six digits) and simplicity of the static value, the value can be easily cracked through brute force attacks. For the other apps, which generate new OTP values, we still regard the PRNGs of these apps as vulnerable because a sufficient time window is left for attackers to retrieve the OTP value. 
We further compared the OTP values generated for the additional account with the original OTP values. All the apps returned a different OTP value for a different account. Therefore, we inferred that the generated OTP values are bound to each user account uniquely.

\subsubsection{Rule 2: Do not generate OTP values in specific patterns}
Among 399 vulnerable apps, 354 apps (88.7\%) are categorized as apps using OTP values in a specific pattern. Once the pattern is recognized, an attacker can launch replay attacks to access the victim's account.

\noindent
\textbf{Rule 2-1: Do not generate a repeated sequence of OTP values.}
\sysname detected 162 (40.6\%) apps that produce the OTP values violating Rule 2-1. In our current implementation of \sysname, we only considered the PRNGs using MT19937 because it is widely used as a default randomness algorithm in several programming languages such as Python and PHP. Therefore, we set $N=624$ to test apps because MT19937 and its variants produce PRNs at a fixed period length (i.e., 624) (see Section~\ref{sub:algo}). Each of their PRNGs generated a pseudo-random stream with 624 unique PRNs and then repeated the same stream. 
Apparently, the PRNGs of these apps use MT19937 as the randomness algorithm. To verify our result, we further executed a seed recovery tool, \tool{untwister}\footnote{untwister: https://github.com/altf4/untwister}, to reverse the PRNG and obtain its original seed. With enough PRNs inputs, \tool{untwister} successfully obtained all their seeds with 100\% confidential.

In this paper, we only inferred whether the randomness algorithm of Mersenne Twister (MT) is implemented in the PRNG. \sysname can be extended easily when we exploited the fix period length from the other randomness algorithms and then \sysname can help with the OTP randomness analysis.

\noindent
\textbf{Rule 2-2: Do not repeat each distinct OTP value $n$ times}
\sysname detected 67 (16.8\%) apps that produce the OTP values violating Rule 2-2. That is, the detected apps iteratively generated the same OTP value $n$ times. According to our manual inspection, we found that 39 apps repeated each OTP value twice, and 27 apps repeated each value three times. Only one app kept repeating the OTP value for five times.

Since there is no randomness algorithm that can generate a value for $n$ consecutive times and then renew the value. We assume that such a scheme is implemented intentionally and the OTP value are stored and being reused when requested. Therefore, we are curious: 1) how long the OTP value will be stored; 2) whether the OTP value changes after being consumed.
Firstly, we conducted three experiments to test how long the OTP value will be stored. Originally, \sysname requested each OTP value after only one minute. We then separately set the waiting period between two login requests as two minutes, 20 minutes and an hour and these OTP values were not consumed. Referring to the previous analysis result, all vulnerable apps repeated the OTP values for less than five times; hence we only sent six login requests in this experiment.
Surprisingly, the results showed that 44 apps renewed their OTP values within 20 minutes and six apps generated new OTP values in an hour. For the rest 17 apps, they did not update the OTP values unless the values were consumed.

Afterwards, we ran three similar experiments, in which the waiting periods were set as the same, but we consumed the OTP value after received. The result demonstrated that 62 apps updated their OTP values immediately after the values were consumed (within one minute). For the rest five apps, one of them updates the OTP value within 20 minutes and the other four apps generated new OTP values in an hour.

\noindent
\textbf{Rule 2-3: Do not generate OTP values with predictable binary representations.}
\sysname detected 125 (31.3\%) apps that produce OTP values violating Rule 2-3. That is, we can predict the OTP values generated from those apps by converting them to binary formats. Among these vulnerable apps, \sysname specifically identified 35 apps generating OTP values with certain shift patterns. Through our manual inspection, six apps iteratively shift all binary digits in anticlockwise direction.
As an example, an app generated a sequence of OTP values as $<081642, 032213, 064426, \dots>$; then, when the numbers are converted from the decimal format into the binary one, the first digit of the current value always appears in the last digit of the next OTP value, i.e., $<$\underline{\textbf{1}}0011111011101010, 0011111011101010\underline{\textbf{1}}, 011111011101010\underline{\textbf{1}}0, \dots$>$.
Instead of shifting the binary digits iteratively, 12 apps add either `1' or `0' at the end of the binary sequence. We also found 17 apps that use similar substitution operations on the digits in other positions.

Besides, \sysname also exploited 90 vulnerable apps that generate OTP values with ``odd-even'' patterns. Within these apps, \sysname discovered eight apps that only generate even OTP values. We can only infer that their developers might implement the randomness algorithm of LFib, MT, or WELL. However, it is difficult for us to identify the use of a specific algorithm in detail from these OTP sequences without accessing the source code of each PRNG.

\subsubsection{Rule 3: Do not use a constant predictable seed to initialize a randomness function}
\sysname detected 4 (0.01\%) apps that produce the OTP values violating Rule 3 -- the PRNGs of three apps are written in C/C++; the PRNG of the other app is written in PHP.
When calling the randomness function \func{rand()}, three apps use a constant \func{srand(1)} to seed \func{rand()}. For the two apps whose PRNGs are written in C/C++, they generate exactly the two same pseudo-random streams. 
Therefore, attackers can rebuild the PRNG to extract the sequence of OTP values. Surprisingly, we identified one app utilizing the timestamp \func{srand(time(NULL))} to seed \func{rand()}. 

\subsection{Insights}
We manually inspected all the analysis results and gained some insights that we report below. 

\begin{table}[!ht]
\centering
\setlength{\abovecaptionskip}{-1pt}
\setlength{\belowcaptionskip}{-100pt}
\caption{\# of vulnerable apps collected from two stores.}

\resizebox{1.0\linewidth}{!}{
\begin{tabular}{ll|c|c|c|c}
\toprule[1.5pt]
\multicolumn{2}{c|}{Violated Rules} & \multicolumn{2}{c|}{\# of \tool{Google Play} Apps} & \multicolumn{2}{c}{\# of \tool{Tencent Myapp} Apps}\tabularnewline

\cline{3-6}
 & & 1-factor & 2-factor & 1-factor & 2-factor\tabularnewline

\midrule[1pt]
\multicolumn{2}{l|}{Rule 1} & -- & 0 & 11 & 30 \tabularnewline
\hline 
\multirow{4}{*}{Rule 2} & Rule 2-1 & -- & 73 & 79 & 10 \tabularnewline
\cline{2-6} 
 & Rule 2-2 & -- & 48 & 10 & 9 \tabularnewline
\cline{2-6} 
 & Rule 2-3 & -- & 15 & 94 & 16 \tabularnewline
\cline{2-6} 
 & Total & -- & 136 & 183 & 35 \tabularnewline
\hline 
\multicolumn{2}{l|}{Rule 3} & -- & 1 & -- & 3 \tabularnewline
\hline
\multicolumn{2}{l|}{Total} & -- & 137 (out of 214) & 194 (out of 1,808) & 68 (out of 1,808)\tabularnewline
\bottomrule[1.5pt]
\end{tabular}
}

\label{table:store}

\end{table}

\noindent
\textbf{App Store Comparison.} As the apps are downloaded from \tool{Google Play} and \tool{Tencent MyApp}, we now discuss the security of the apps collected from each store. Table~\ref{table:store} presents the number of vulnerable apps collected from each store. Among the 214 apps that were downloaded from \tool{Google Play}, \sysname detected a randomness vulnerability in 137 (64\%) apps while it detected a randomness vulnerability in 262 (14.5\%) apps out of 1,808 apps downloaded from \tool{Tencent MyApp}. At first glance, the \tool{Google Play} apps would be more vulnerable compared with the \tool{Tencent MyApp} apps. However, all the \tool{Google Play} apps use a two-factor authentication scheme, which means that the additional user credential information is needed to attack such an app from \tool{Google Play}. In contrast, 194 (74\%) out of the 262 vulnerable \tool{Tencent MyApp} apps use the OTP authentication alone without any additional security mechanisms, leading to insecure authentication against guessing attacks and replay attacks.

Furthermore, we specifically examine the difference between both app stores in terms of each randomness rule. For Rule 1, \sysname discovered 41 vulnerable apps from \tool{Tencent MyApp} only. For Rule 2, we can see a different distribution between two stores -- among the 136 vulnerable \tool{Google Play} apps, Rule 2-1 (73 violations), Rule 2-2 (48 violations), and Rule 2-3 (15 violations) are most frequently found in order while among the 218 vulnerable \tool{Tencent MyApp} apps, Rule 2-3 (110 violations), Rule 2-1 (89 violations), and Rule 2-2 (19 violations) are most frequently found in order. For Rule 3, there are only a few vulnerable apps (1 in \tool{Google Play} and 3 in \tool{Tencent MyApp}) commonly in both app stores. The use of the Chi-square homongeneity test~\cite{alexander1989statistical} revealed that there is a significant difference in the ratios of violated rules between two app stores ($p < 0.05$, $\chi^2=169.827$, $df = 4$).

\noindent
\textbf{App Category Security.}
As well as analyzing from which app stores the apps are from, we additionally analyzed the vulnerable apps by categories. 
We discovered that the top three most vulnerable categories are Video Players \& Editors, Entertainment, and Music \& Audio. The numbers of vulnerable apps from those categories are 182 (45.61\%), 71 (17.80\%), and 34 (8.52\%), respectively.
Although these apps might not be as crucial as apps such as Finance and Social, these apps also store private and sensitive user data. For example, the user often use the Video Players \& Editors to edit her private video clips and save them as drafts in the account. The video clips in the draft box are stored in the cloud storage sometimes instead of the user's local storage on her mobile phone. From such Video Players \& Editors apps, the attacker can easily access the user's account and steal the stored private video clips. 

\subsection{Limitations}
\label{sub:limitation}

\noindent
\textbf{Uncertainty about Involved Parameters.}
Referring to the analyses in Section~\ref{sec:pre}, \sysname can only determine whether the PRNG of an app is using a known cryptographically insecure algorithm to generate OTP values by analyzing their patterns. Nevertheless, in some cases, it is unable to decide what PRNG parameters such as the seeds are used for the PRNG implementation on the server-side. To overcome this limitation, we should gather more datasets with various parameter values for PRNGs.

\noindent
\textbf{Limited Analysis Scope.}
\sysname detects all violations of randomness rules based on the code samples that are collected from GitHub and Stack Overflow. Since we only focus on specific randomness algorithms and randomness functions based on our codebase, the detection ability of \sysname is inherently limited to specific algorithms and functions. However, in practice, there are various PRNG algorithms and implementations that we do not cover in this paper. For example, \tool{Clipperz}\footnote{Clipperz: https://github.com/clipperz/javascript-crypto-library} is a Javascript crypto library that we did not consider yet. To evaluate the randomness and predictability of new and unseen PRNGs, we need to introduce more robust statistical tests in the existing PRNG guidelines (e.g., NIST SP 800-22~\cite{Bassham10:randomness}).

\noindent
\textbf{Failed Apps Analysis.}
\sysname failed to decompile or analyze 2,416 Android apps. By manually inspecting these apps, we found that 889 apps use code packing to defend against decompilation. Since their ``class'' files are encrypted, \tool{JEB} cannot decrypt the file to extract their source code. In addition, 412 apps have obfuscated code that cannot be analyzed. For 1,115 apps, \sysname failed to send login requests because a runtime error occurred while sending request messages.

\section{Related Work}
\label{sec:related}

\noindent
\textbf{Analysis of PRNG.}
Several vulnerability analysis studies have been conducted on PRNG. 
Most of them focus on the PRNG in Linux because it is included in the kernel of all Linux distributions and widely used in many security-related applications and protocols~\cite{gutterman2006analysis}~\cite{dodis2013security}~\cite{everspaugh2014not}~\cite{ristenpart2010good}.
As the Linux PRNG is open source, Dodis \emph{et al.}~\cite{dodis2013security} and Gutterman \emph{et al.}~\cite{gutterman2006analysis} assessed its security, relying on the code analysis. Dodis \emph{et al.} proposed a new formal security model for PRNGs , which encompassed all the proposed security notions. Then, they evaluated the security of the two Linux PRNGs, \func{/dev/random} and \func{/dev/urandom}, and proved that these PRNGs are not robust and do not accumulate entropy properly.
Aside from analyzing the source code, Gutterman \emph{et al.}~\cite{gutterman2006analysis} combined static reverse engineering of the source code with dynamic tracing to learn the operation of the PRNG. 

Apart from Linux PRNG, OpenSSL is another important application for PRNG analysis~\cite{kim2013predictability}~\cite{oak2019poster}. Kim \emph{et al.}~\cite{kim2013predictability} investigated the Android OpenSSL PRNG. Similarly, they conducted a code analysis to analyze the Android OpenSSL architecture. Starting from the initialization, they found that every SSL application generates random data from the same initial state. Thus, attackers can recover the state of the OpenSSL PRNG from any apps.

However, the above approaches cannot be applied to analyze the PRNG for OTP authentication because the app PRNG is not open source and typically implemented on the server-side. Therefore, we performed a black-box analysis to detect what algorithm might be implemented and then verified our assumption through dynamic analysis to demonstrate vulnerabilities in the generated stream of OTP values.

Similar to our work, Argyros \emph{et al.}~\cite{argyros2012forgot} focused on the predictability of password reset token and exploited PHP randomness generators. By analyzing the randomness functions supported by PHP, they launched attacks from the timestamp and the seed aspects, respectively. Without the specific target of the programming language, our analyzing objects are more general.

\noindent
\textbf{OTP Authentication.}
We categorize the previous work on the security of OTP authentication into two groups: vulnerability analysis and security enhancement. 

Mulliner \emph{et al.}~\cite{mulliner2013sms} analyzed SMS OTP authentication from the perspectives of access control and involved parties. They explored the potential weaknesses and introduced attacks to exploit them. Several basic countermeasures were given to defend against the attacks. Instead of analyzing the general OTP authentication, Yoo \emph{et al.}~\cite{yoo2015case} and \tool{AUTH-EYE}~\cite{ma2019empirical} analyzed specific scenarios, i.e., internet banking services, and Android OTP authentication, respectively. Yoo \emph{et al.} investigated the security measures in internet banking and discovered a novel type of attack for hijacking the implemented OTP system. \tool{AUTH-EYE} presented six OTP rules for the implementation of secure OTP authentication applications. It checked whether OTP authentication violated any of these rules and determined what implementation error existed in the authentication scheme.

Several approaches have been proposed to provide secure OTP authentication methods. Das \emph{et al.}~\cite{das2017secure} combined the image with numeric values to ensure the unpredictable feature for each OTP. They selected a pseudo-random value as the first part and then randomly selected pixels of user biometric image as the second part. A few studies on OTP authentication schemes focus on specific scenarios. E.g., Jeong \emph{et al.}~\cite{jeong2008integrated} designed an OTP authentication scheme for home networks with low computation, and Rifa \emph{et al.}~\cite{rifa2009secure} designed an OTP authentication scheme for e-banking by combining symmetric cryptography with a hardware security module.

Nonetheless, none of the above approaches assesses the randomness of the generated OTP values. Our work is the first to systemically analyze the randomness of OTP values used in real-world Android apps.

\section{Conclusion}
\label{sec:conclusion}
In this paper, we analyzed randomness algorithms and functions and then introduced three randomness rules that must be followed in the implementation of cryptographically secure pseudo-random number generators (PRNGs). To analyze the PRNG implementations, we designed an automated analysis tool, \sysname, to investigate the randomness of OTP values generated by PRNGs. Without accessing the implementation source code, we collected PRNG implementation codes shared by other developers to learn the most common implementations. Finally, we assessed 6,431 real-world Android apps against the randomness rules and detected 399 apps that generate vulnerable OTP values. Our findings demonstrated that a significant number of existing OTP-based services can be easily attacked in practice, and thus we need to promptly fix those apps to protect users. Perhaps the use of naive PRNGs for OTP authentication is another example of ``security theater,'' which tackles the feeling but not the reality.

\section*{Acknowledgement}
This work was supported by the start-up grant of the School of Information Technology and Electronic Engineering, The University of Queensland, and partially supported by the National Natural Science Foundation of China (Grant No.62002222), the National Key Research and Development Program of China (Grant No.2020AAA0107800).


\end{document}